\documentclass{ws-mpla}
\usepackage[super]{cite}
\usepackage{graphicx}

\begin{document}

\markboth{Q. Le Thien, D. E. Krause}
{Unitarily Inequivalent Vacua and Long-Range Forces: 
Phenomenology with Scalar Boson Mass-Shift}

\catchline{}{}{}{}{}

\title{Unitarily Inequivalent Vacua and Long-Range Forces:\\
Phenomenology with Scalar Boson Mass-Shift}

\author{\footnotesize Quan Le Thien}

\address{Physics Department, Wabash College, Crawfordsville, IN 47933, USA}

\author{Dennis E. Krause}

\address{Physics Department, Wabash College, Crawfordsville, IN 47933, USA \\
	Department of Physics  and Astronomy, Purdue University, West Lafayette, IN 47907, USA}
\maketitle


\begin{abstract}
We explore the impact of a sudden shift in the mass of a scalar boson field on long-range forces mediated by this field under the framework of unitarily inequivalent vacua. Since the search for new long-range forces is an active experimental area probing  physics beyond the Standard Model, the consequence of a non-trivial vacuum state of a scalar boson on these experiments is elucidated. We show that while the mass shift affects the one-boson exchange potential, the Casimir force remains only dependent on the vacuum state.

\end{abstract}

\ccode{PACS Nos.: 11.90.+t.}

\section{Introduction}

Even after the discovery of the Higgs boson, the mechanisms for generating mass remains an active research area relating  neutrinos, dark matter, dark energy, and other Beyond Standard Model (BSM) physics. One of the Higgs boson's central features is its metastable vacuum, which is currently believed to decay eventually \cite{Andreassen}. A catastrophic consequence of this decay is to turn off the Higgs mechanism for all elementary particles, effectively rendering all of them to be massless. Cosmological aspects relating to the phenomenon of vacuum decay have been readily explored \cite{Markkanen}. From a phenomenological perspective, two immediate interesting questions arise: firstly, whether this phenomenon of a change in the mass-generating mechanism has happened before, and secondly, whether we can know about these past events from experiments conducted in our time. Hence, the purpose of this paper is to study the consequence of such an effective mass-shift which may be the result of some unknown exotic Higgs physics in the distant past. Our result indeed shows that observers after the transition should still be able to observe artifacts of physics before the transition. Particularly, we examine effects of the mass-shift on the vacuum structure of a scalar boson in a toy model and its associated physics probed by low-energy long-range force experiments. 

To simplify the phenomenology, we assume that the change in the (Higgs-like) mass-generating mechanism can be treated as abrupt in the sense that it only changes the mass of the scalar boson in the Lagrangian without affecting the current values and states of the scalar boson field. Then, we will see that such a mass-shift only changes the original representation of the field, when canonically quantized, into one of a new mass, while the vacuum state remains the same. This mismatch between the new representation of the field and the old vacuum state necessitates the use of a Bogolyubov transformation to connect creation and annihilation operators belonged to unitarily inequivalent vacua of different masses. In fact, such a mismatch has emerged recently in the context of mixing fermions \cite{bv fermion, tureanu prd neutron} and bosons \cite{bv boson} in quantum field theory (QFT). Although the freedom to choose the physical mass representation\cite{bv boson,bv fermion} is precisely what this approach is most criticized for in Ref.~\refcite{Giunti}, it is important to recognize that, phenomenologically speaking, the universe we are living in might indeed favor one representation over others. Our results indicate that long-range forces and potentials are observables which reveal the difference between physical representations of the field and vacuum should it exist. While the mass-shift model considered here is significantly different from that used to describe particle mixing  \cite{bv boson,bv fermion}, due to the similarities in the employment of unitarily inequivalent vacua, our work might offer some insights into the Casimir effect for mixed boson fields \cite{bv casimir}.    

While  long-range forces and potentials are cornerstones of classical physics, under the modern quantum field-theoretic framework they arise as a consequence of quantum processes. Hence, by studying these long-range forces one can infer fundamental properties of particles involved and their associated physics. Certainly, this is not a new idea, various searches for new physics are studying long-range forces and potentials in a plethora of systems ranging from atoms and molecules \cite{review new physics atoms} to astrophysical objects \cite{review new physics macroscopic}. However, most of these existing approaches focus on new physics manifested as additional coupling terms to the effective Lagrangian of the Standard Model (SM). In this paper, we study a novel scenario where BSM physics originates from the non-trivial structure of the vacuum state which results in deviations from the usual Yukawa-type One-Boson Exchange Potential (OBEP) and the well-known Casimir force. Our results arise out of the interesting fact that the OBEP and the Casimir force arise from different physical processes. The OBEP involves a virtual-particle exchange that depends on the coupling constant governing the interaction of the boson and fermion, while the Casimir force is a consequence of the change in the vacuum energy due to boundary conditions.\cite{Milonni,Casimir,Bordag}  We will see that in our model the mass-shift manifests different phenomena in these two long-range interactions, which suggests that different types of force experiments may play complementary roles in investigating these types of phenomena.

Unitarily inequivalent representations of the canonical commutation relations have long been recognized as the dichotomous distinction between quantum mechanics and QFT \cite{Umezawa book,Blasone book,Grib}, because of the failure of Stone-von Neumann uniqueness theorem \cite{Neumann} when applied to the latter. The existence of these representations lead inevitably to an infinite number of inequivalent physical realizations of the described dynamics. Despite being identified early in the development of QFT, the physical interpretation of this issue still remains poorly understood in the context of elementary particle physics \cite{Grib, Miransky book}. Nonetheless, this feature of QFT has been shown to be absolutely necessary in order to describe various physical scenarios such as macroscopic quantum systems in condensed matter physics \cite{Miransky book,Blasone book}, spontaneous symmetry breakdowns \cite{Umezawa book,Miransky book}, Thermo Field Dynamics (TFD) \cite{Takahashi, Umezawa book, Das} and QFT on curved spacetime \cite{Parker}. Hence, it is possible that methods of unitarily inequivalent representations are physically meaningful to the fundamental laws of physics. Typically, modern practice of particle physics ties the existence of these representations together with the famous Haag's theorem \cite{Haag} and thus ignores their relevance. In this paper, we show that unitarily inequivalent representations help to describe, within certain limits, the phenomenon of mass-shift as a result of a change in the mass-generating mechanisms. Our consideration here hopefully will serve as an explicit example of how unitarily inequivalent representations can be used to study exotic phenomena in potential new physics.

The rest of this paper is organized as follows. The next two sections are dedicated to describing our toy model with a scalar field where issues with the field representation and unitarily inequivalent vacua are motivated. Section \ref{condensate} studies the most obvious consequence of our toy model, namely how it induces a condensate after the mass-shift. Then Sections \ref{OBEP section} and \ref{Casimir section} study the effects of the mass-shift on two different types of long-range forces, the OBEP and Casimir force, respectively. Overall, we discuss connections of this toy model to relevant physical scenarios to the SM. Lastly, we  note that during the process of finalizing our calculation in Section \ref{Casimir section}, two models involving similar unitarily inequivalent vacua have been proposed and studied in the context of soliton field theory \cite{Zhou} and weakly interacting continuum field theory \cite{Cotler}. Furthermore, such unitarily inequivalent vacua were also studied by Miransky \cite{Miransky book}.

\section{Mass-Shift: Toy Model with Scalar Boson}

We begin by considering the free Lagrangian density for a real scalar field $\phi(x)$ with mass $m_1$ in a four-dimensional Minkowski spacetime ($\hbar = c = 1$):
\begin{equation}
\mathcal{L}^{\rm free}_{1} = \frac{1}{2}  \partial_\mu \phi\left(x\right) \partial^\mu \phi \left(x\right) - \frac{1}{2} m_1^2 \phi^2\left(x\right). 
\label{lagrangian density 1}
\end{equation}
Due to a phase transition in a Higgs mechanism caused by vacuum decay \cite{Andreassen} or some other unknown mass generating mechanism, the initial mass $m_1$ of the field is abruptly changed to $m_2$, so the new free Lagrangian density now reads
\begin{equation}
\mathcal{L}^{\rm free}_{2} = \frac{1}{2}  \partial_\mu \phi\left(x\right) \partial^\mu \phi\left(x\right) - \frac{1}{2} m_2^2 \phi^2\left(x\right). 
\label{lagrangian density 2}
\end{equation}
These two Lagrangian densities in Eqs.~(\ref{lagrangian density 1}) and (\ref{lagrangian density 2}) correspondingly lead to two different Hamiltonians before and after the transition,
\begin{equation}
H_i = \int d^3 r \left[ \frac{1}{2} \Pi^2\left(x\right) + \frac{1}{2} \nabla^2 \phi\left(x\right) + \frac{1}{2} m_i^2 \phi^2\left(x\right) \right],
\label{hamitonian i}
\end{equation}
where $i = 1,2$, which have distinct excitation spectra due to the different masses involved. Using the canonical quantization procedure in the Schr\"odinger picture, we can expand the fields as
\begin{equation}
\phi(\vec{r}) = \left\{ 
\begin{array}{ll} 
\displaystyle \phi_{1}(\vec{r}) \equiv \frac{1}{\sqrt{V}} \sum_{\vec{k}} \frac{1}{\sqrt{2\omega_1}} \left[  a_1^{}  + a_{-,1}^\dagger  \right] e^{i \vec{k} \cdot \vec{r}}, & \mbox{before the transition,}
\\ 
&\\
\displaystyle \phi_{2}(\vec{r}) \equiv \frac{1}{\sqrt{V}} \sum_{\vec{k}} \frac{1}{\sqrt{2\omega_2}} \left[  a_2^{}  + a_{-,2}^\dagger  \right] e^{i \vec{k} \cdot \vec{r}}, & \mbox{after the transition,}
\end{array}
\right.
\label{field expansion mass 2}
\end{equation}
where $\phi_{1}(\vec{r})$ and $\phi_{2}(\vec{r})$ represent the field expansions before and after the transition.  Here
\begin{eqnarray}
a^\dagger_{-,i} &\equiv & a^\dagger_{-\vec{k},i}, \\
a_i&\equiv& a_{\vec{k},i}, \\
\omega_i &\equiv& \sqrt{k^2 + m_i^2},
\end{eqnarray}
where $a^\dagger_{-,i}$ and $a_{i}$  are the creation and annihilation operators, respectively, and $\omega_i$ is the energy of the field elementary excitations. The corresponding vacuum states $|0\rangle_{1}$ and $|0\rangle_{2}$ are defined by
\begin{equation}
a_{i}|0\rangle_{i} \equiv 0.
\end{equation}
 
Now we are presented with a tension. If we are an observer after the transition from $H_1$ to $H_2$, we should observe elementary excitations of the physical representation of the field expansion $\phi_{2}\left(x\right)$ to be eigenstates of $H_2$ with dynamical mass is $m_2$. The reason why we call $\phi_{2}(\vec{r})$ the physical representation of the field expansion is because it diagonalizes the Hamiltonian $H_2$ in Eq.~(\ref{hamitonian i}). A similar procedure has been used to determine the physical representation of neutron field in neutron-antineutron oscillation \cite{tureanu prd neutron}. It is then tempting to draw the conclusion that the physical vacuum of this system is the eigenstate $|0\rangle_2$ of $a_2$, because it is the eigenstate corresponding to the lowest eigenenergy of $H_2$ in Eq.~(\ref{hamitonian i}). However, this cannot be true, since the same argument can be applied to observers before the transition who will arrive at $|0\rangle_1$ being the physical vacuum state. In this paper, we will resolve this tension by assuming that the mass-shift from $H_1$ to $H_2$ is considered to be abrupt, as defined earlier, thus the chronological order of the two Lagrangians determines the physical vacuum state of the system. Particularly, the vacuum state $|0\rangle_1$ before the transition remains the physical vacuum for this system even after the transition. 

One immediate criticism of our approach might come from the fact that $|0\rangle_1$ is no longer the eigenstate corresponding to the lowest eigenenergy of $H_2$ after the transition. While a quantum system is generally expected to be at its ground state to minimize its energy, such a condition is not necessary when the system experiences symmetry breaking. In our model, such mechanism is provided by the mass-shift. Furthermore, not only does the new zero-particle state differs from the physical vacuum, but the entire new Hilbert space built upon this zero-particle state does not contain the physical vacuum\cite{Grib,Umezawa book,Blasone book,Miransky book}. Hence, it is clear that the physics here necessitates the use of unitarily inequivalent vacua. These subtleties has been encountered in recent studies of neutron-antineutron conversion \cite{tureanu prd neutron} and particle mixing \cite{bv boson,bv fermion}. Particularly, issues with the vacuum state in our mass-shift model share many theoretical features with ones from the well-known problem of particle creation in expanding universes \cite{Parker}.

\section{Fields and Vacua: Representations}

In quantum field theory, the observed dynamics is determined by two ingredients: the states and fields. For observers after the transition, $\phi(x)$ will take the physical representation $\phi_{2}(x)$, thus the time-evolution of  $\phi_{2}(x)$ is trivially governed by $H_2$. Nonetheless, the states are indeed built upon the vacuum state $|0\rangle_1$. Hence, in order to perform calculations, we need to relate the physical creation and annihilation operators of $H_2$ with the corresponding of $H_1$. This can be achieved by matching the two representations of the field $\phi(\vec{r})$ and its conjugate field $\Pi(\vec{r})$ before and after the transition:
\begin{eqnarray}
\label{phi matching} 
\phi(\vec{r}) = \frac{1}{\sqrt{V}} \sum_{\vec{k}} \frac{1}{\sqrt{2\omega_1}} \left[  a_{-,1}^\dagger + a_1^{}  \right] e^{i \vec{k} \cdot \vec{r}}  = \frac{1}{\sqrt{V}} \sum_{\vec{k}} \frac{1}{\sqrt{2\omega_2}} \left[ a_{-,2}^\dagger +   a_2^{}  \right] e^{i \vec{k} \cdot \vec{r}}, \\
\label{pi matching}
\Pi(\vec{r}) = \frac{1}{\sqrt{V}} \sum_{\vec{k}} \sqrt{\frac{\omega_1}{2}} \left[  a_{-,1}^\dagger - a_1^{}  \right] e^{i \vec{k} \cdot \vec{r}} =  \frac{1}{\sqrt{V}} \sum_{\vec{k}} \sqrt{\frac{\omega_2}{2}} \left[   a_{-,2}^\dagger - a_2^{}   \right] e^{i \vec{k} \cdot \vec{r}}.
\end{eqnarray}
It is important to note that in writing down Eq. (\ref{phi matching}) and (\ref{pi matching}), we are implementing the assumption that the fields $\phi(x)$ and $\Pi(x)$ are unaffected by the mass-shift. To proceed, we compare the mode expansion in LHS and RHS of Eq.~(\ref{phi matching}) and (\ref{pi matching}) to obtain
\begin{equation}
\left(\begin{array}{cc}
1/\sqrt{2\omega_1} & 1/\sqrt{2\omega_1}\\
\sqrt{\omega_1/2} & - \sqrt{\omega_1/2}
\end{array}\right) \left(\begin{array}{c}
a_{-,1}^\dagger\\
a_1
\end{array}\right) =  \left(\begin{array}{cc}
1/\sqrt{2\omega_2} & 1/\sqrt{2\omega_2}\\
\sqrt{\omega_2/2} & - \sqrt{\omega_2/2}
\end{array}\right) \left(\begin{array}{c}
a_{-,2}^\dagger\\
a_2
\end{array}\right).
\end{equation} 
Then, we can easily obtain the desired Bogolyubov transformation between different representations of the canonical operators as follows,
\begin{equation}
\left(\begin{array}{c}
a_{-,2}^\dagger\\
a_2
\end{array}\right) = \left( \begin{array}{cc}
u & v\\
v & u\\
\end{array} \right) \left(\begin{array}{c}
a_{-,1}^\dagger\\
a_1
\end{array}\right),
\label{bogolyubov transformation}
\end{equation}
where $u$ and $v$ are given by
\begin{eqnarray}
u &=& \frac{1}{2} \left(\sqrt{\frac{\omega_1}{\omega_2}} + \sqrt{\frac{\omega_2}{\omega_1}} \right), \\
v &=& \frac{1}{2} \left(\sqrt{\frac{\omega_2}{\omega_1}} - \sqrt{\frac{\omega_1}{\omega_2}} \right).
\end{eqnarray}
It is clear from Eq.~(\ref{bogolyubov transformation}) that the Bogolyubov transformation employed here only mixes canonical operators with the same momenta, thus it can be considered in many ways as the transformation of quantum mechanical operators with a finite number of degrees of freedom. It is also pointed out that what we are doing in Eq.~(\ref{phi matching}) and (\ref{pi matching}) is to rescale the mass, effectively yielding a dilation \cite{Zhou}. Furthermore, this mass rescaling is the precise analog of the varying metric in expanding universes \cite{Parker}. 

An important remark follows from the analog with the particle creation in expanding universes as noted above is the structure of the vacuum. Since observers' notion of a particle is dictated by the representation of the creation and annihilation operators, although the vacuum state does not change, observers after the mass-shift should see the vacuum as filled with pairs of particles \cite{Parker}, because the definition of a particle has changed. Particularly, we can write down the vacuum state $|0\rangle_1$ in the new representation
\begin{equation}
|0\rangle_1 = \prod_{\vec{k}} \left[ c_0(\vec{k}) |0\rangle_2 + c_2(\vec{k}) |1_{\vec{k}} 1_{-\vec{k}} \rangle_2 + c_4(\vec{k}) |2_{\vec{k}} 2_{-\vec{k}} \rangle_2 + ... \right],
\label{vacuum structure}
\end{equation} 
where $c_i(\vec{k})$ can be determined from the Bogolyubov transformation in Eq.~(\ref{bogolyubov transformation}). Details of the vacuum structure have already been worked out explicitly in Fock space \cite{Miransky book} and in the  Schr\"odinger representation \cite{Zhou}. They are not included here since they are not required for subsequent calculations in this paper. Nonetheless, Eq.~(\ref{vacuum structure}) indicates clearly that $|0\rangle_1$ evolved non-trivially in time by $H_2$. This means that even though our Lagrangian density after the mass-shift $\mathcal{L}_2$ is Lorentz-invariant, the observed dynamics is not because the vacuum state is time-dependent and thus no longer Lorentz-invariant.

\section{Condensate Density}
\label{condensate}

As pointed out in the previous sections, because our system bears many resemblances with the process of particle creation in expanding universes \cite{Parker}, we will investigate how the mass-shift can lead to an observable condensate density. Let us consider the particle density
\begin{equation}
\rho = \frac{\langle \hat{N} \rangle }{V},
\label{condensate density}
\end{equation}
where $V$ is the volume of the system, in this case being one of the current universe, and $\langle \hat{N} \rangle$ is the vacuum expectation value (vev) of the number operator. In this case, the physical representation of the number operator belongs to the one of mass $m_2$, while the vacuum state belongs to that of $m_1$: 
\begin{equation}
\langle \hat{N} \rangle = {}_1\langle 0| \hat{N}_2 |0\rangle_1.
\label{N2 vev wrt vac 1}
\end{equation}
where the number operator $\hat{N}_2$ is given by
\begin{equation}
\hat{N}_2 = \sum_{\vec{k}} a^\dagger_2 a_2. 
\label{N2}
\end{equation}

In order to calculate compute $\langle \hat{N} \rangle$ in Eq.~(\ref{N2 vev wrt vac 1}), we substitute the Bogolyubov transformation in Eq.~(\ref{bogolyubov transformation}) into Eq.~(\ref{N2}) to obtain
\begin{eqnarray}
\nonumber
\langle \hat{N} \rangle = {}_1\langle 0| \hat{N}_2 |0\rangle_1 &=& {}_1\langle 0|\sum_{\vec{k}} \left( u\ a_{1}^\dagger + v\ a_{-,1}^{} \right) \left( v\ a_{-,1}^\dagger + u\ a_{1}^{} \right) |0\rangle_1 \\
\nonumber
&=& {}_1\langle 0|\sum_{\vec{k}} \left[uv \left( a_{1}^\dagger a_{-,1}^\dagger + a_{-,1} a_1^{}  \right) + u^2 a_1^\dagger a_1^{} + v^2 a_{-,1} a_{-,1}^\dagger  \right] |0\rangle_1 \\
\label{N2 vev wrt vac 1 result}
&=& \sum_{\vec{k}}  v^2.
\end{eqnarray}
To proceed, we apply the continuum limit,
\begin{equation}
\sum_{\vec{k}} \rightarrow \frac{V}{\left(2\pi\right)^3} \int d^3k,
\label{continuum limit}
\end{equation}
to the sum in Eq.~(\ref{N2 vev wrt vac 1 result}), and substitute the result back into the condensate density in Eq.~(\ref{condensate density}) to obtain
\begin{equation}
\rho = \frac{1}{(2\pi)^3} \int d^3k \ \frac{\left(\omega_2 - \omega_1\right)^2}{4\omega_1 \omega_2} \
= \frac{1}{8\pi^2} \int_{0}^{\infty} dk\ k^2 \frac{\left(\omega_2 - \omega_1\right)^2}{\omega_1 \omega_2}.
\label{rho vev wrt vac 1 result}
\end{equation}

\begin{figure}
	\centering
	\includegraphics[width=13cm]{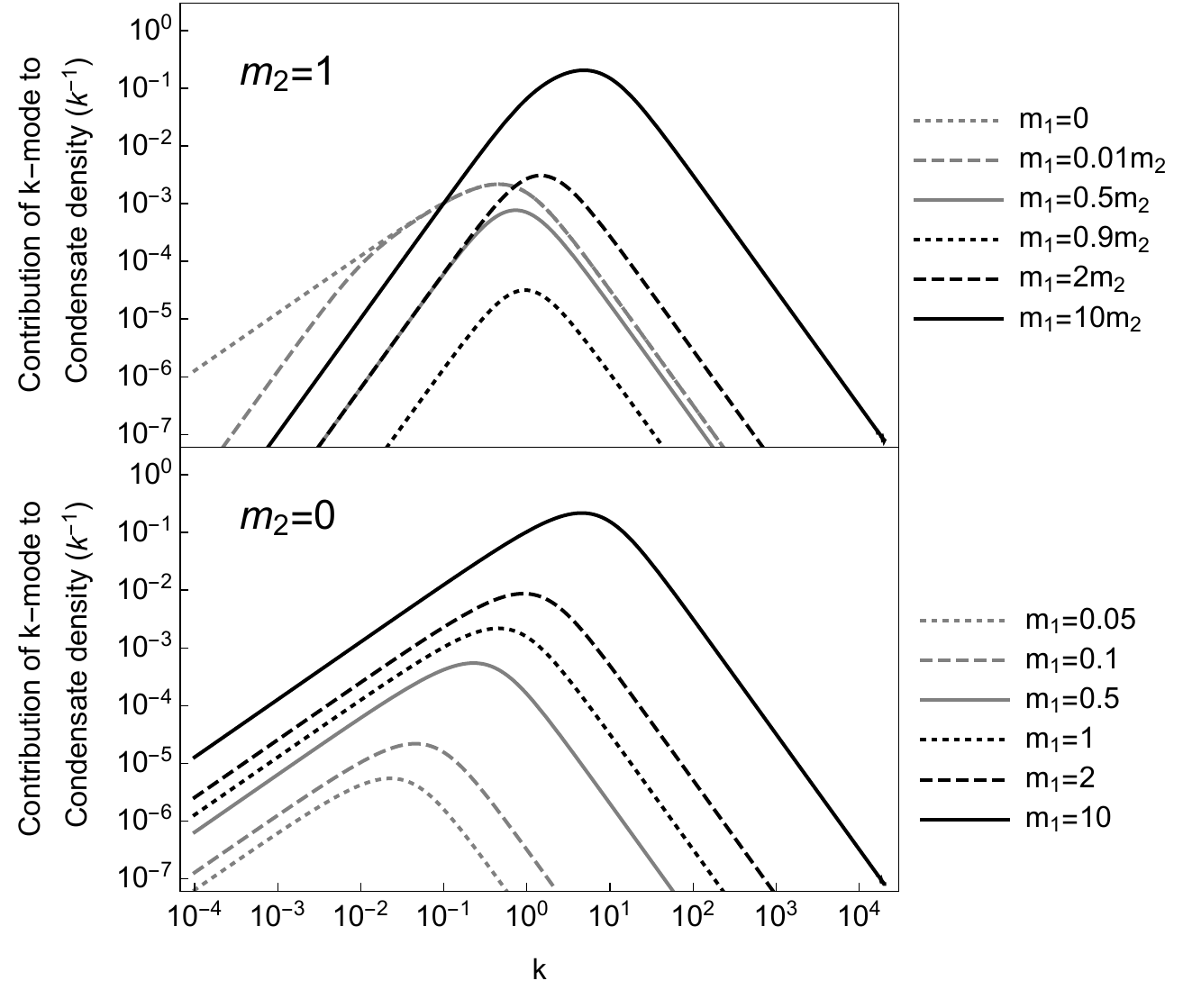} 
	\caption{Plot of contribution of $k$-modes to the condensate density. The upper panel explores the case where the mass after the mass-shift is non-zero, setting $m_2=1$ (in arbitrary units), showing various values for the mass before mass-shift $m_1$: $0$ (dotted gray), $0.01 m_2$ (dashed gray), $0.5m_2$ (solid gray), $0.9m_2$ (dotted black), $2m_2$ (dashed black) and $10m_2$ (solid black). Meanwhile, the lower panel considers the situation where $m_2=0$ for a range of values of $m_1$ (in arbitrary units):  $0.05$ (dotted gray), $0.1$ (dashed gray), $0.5$ (solid gray), $1$ (dotted black), $2$ (dashed black) and $10$ (solid black).}
	\label{fig condensate density contribution massive}
\end{figure}

While the integral in Eq.~(\ref{rho vev wrt vac 1 result}) can be evaluated exactly, the result is not particularly enlightening. Instead, we can extract the contribution of each $k$-mode to the condensate density from Eq.~(\ref{rho vev wrt vac 1 result}) as
\begin{equation}
\rho(k)= \frac{k^2}{8\pi^2} \frac{\left(\omega_2 - \omega_1\right)^2}{\omega_1 \omega_2},
\label{k-mode contribution}
\end{equation}
which has the physical meaning as the number density per unit volume per unit $k$. It is interesting to see in Fig.~\ref{fig condensate density contribution massive} that both the large $k$-modes' (ultrarelativistic) and the small $k$-modes' (nonrelativistic) contributions to the condensate are suppressed out. While the exact expression for the peak $k$-mode in Fig.~\ref{fig condensate density contribution massive} is complicated, when $m_{1} < m_{2}$, the peak position ranges between $k = (1 - 1/\sqrt{2})^{1/2}m_{2} \simeq 0.5412m_{2}$ when $m_{1} = 0$ to $k = m_{2}$ when $m_{1}/m_{2} \rightarrow 1$. Another interesting observation is that both Eq.~(\ref{rho vev wrt vac 1 result}) and (\ref{k-mode contribution}) are symmetric in terms of $m_1$ and $m_2$, which means that the observation of condensate calculated here alone does not allow one to directly determine which mass state the vacuum is currently in.

While the effect of a particle-pairs' condensate, such as one in Eq.~(\ref{vacuum structure}), on collider experiments need to be elucidated further, one implication of the behaviors in Fig.~\ref{fig condensate density contribution massive} is that current high-energy experiments might not have the sufficient sensitivity to probe the presence of this condensate due to the this suppression, if the values of $m_1$ and $m_2$ are either too large or small. Since the vacuum described by Eq.~(\ref{vacuum structure}) is filled with particles, one will expect a non-zero vev of the energy-momentum tensor. Such non-vanishing behavior should leave trails on a cosmological scale via gravitation, acting either as a dark energy or dark matter component. Experiments looking for these large-scale signatures will not encounter similar difficulties as aforementioned high-energy experiments, because the vev of the energy-momentum tensor aggregates over all $k$-modes of the condensate.

\section{One-Boson Exchange Potential with Mass Shift}
\label{OBEP section}

\subsection{Formalism and Remarks}
In this section, we examine an important low-energy consequence of the mass-shift, the One-Boson Exchange Potential (OBEP) arising from a virtual-exchange on the condensate in Eq.~(\ref{vacuum structure}). Let us consider the following interaction Lagrangian density 
\begin{equation}
\mathcal{L}_{\rm int} = - g	 \phi(x) J(x),
\label{L int density}
\end{equation}
where $\phi(x)$ is a real scalar field and $J(x)$ is the external static particles' source density. If $\phi(x)$ is taken to be a real scalar field interacting with two stationary particles with couplings $g_{1}$ and $g_{2}$,  Eq.~(\ref{L int density}) at lowest-order leads to the usual OBEP, also known as the Yukawa potential \cite{Padmanabhan}
\begin{equation}
V_Y (r) = - \frac{g_1 g_2e^{-mr}}{4\pi r},
\label{Yukawa Potential}
\end{equation}
where $m$ is the exchanged boson's mass. However, since $\phi(x)$ experienced a mass-shift as proposed earlier, this result is no longer valid. 

In order to understand the effect of the condensate on the OBEP, let us first quickly review the procedure of obtaining the typical OBEP. In the functional approach, the potential is related to the vacuum-vacuum transition amplitude\cite{Padmanabhan}
\begin{equation}
\langle 0_+ | 0_-\rangle^J = \exp -\frac{1}{2} \int d^4x_1 d^4x_2 \ J(x_1) G(x_2;x_1) J(x_2),
\label{vacuum vacuum amplitude}
\end{equation}
where $G(x_2;x_1)$ is the Green's function of the Klein-Gordon equation subjected to appropriate boundary conditions. Then, the integral is evaluated assuming the static sources and identified as
\begin{equation}
\langle 0_+ | 0_-\rangle^J = \exp\left( -i \int_{t_i}^{\infty} dT \ \mathcal{E}\right),
\label{potential prescription}
\end{equation} 
where $T$ can be interpreted as the total interaction time of the two sources, $t_i$ is the time when the interaction is turned on and $\mathcal{E}$ is proportional to the desired Yukawa potential in Eq.~(\ref{Yukawa Potential}). Although by static sources we mean $J(x)$ is a time-independent function signifying the sources' constant presence from the distant past to the distant future, this is generally not true since external particles should be created after a certain cosmological time-scale. The OBEP is really due to the interaction switched on at $t_i$, which can be understood as when the sources were created. While in deriving Eq.~(\ref{Yukawa Potential}) the limit $t_i\rightarrow-\infty$ is used, in our model, we have to be careful since the mass-shift defines another peculiar point in time. In this work, we will assume that our sources started interacting after the mass-shift, corresponding to the interpretation that test objects involved in the table-top experiments \cite{review new physics atoms,review new physics macroscopic} interact with $\phi(x)$ in the $\phi_2$ representation. While one can reasonably ask what happens to the other scenario where the sources are turned on before the mass-shift, it is hard for us to look for a physical system where such OBEP is involved.

While the standard approach laid out in the previous paragraph is theoretically insightful, the boundary conditions used to obtain the Green's function in Eq.~(\ref{vacuum vacuum amplitude}) are not physically opaque in our mass-shift model. In fact, issues with boundary conditions in general situations involving unitarily inequivalent vacua have been shown to require quite subtle treatments under the functional approach \cite{Blasone path integral qft, Blasone path integral qm}. Hence, instead of using the aforementioned functional technique, we proceed with the more standard diagrammatic approach for calculating potentials from the Lorentz-invariant Feynman amplitudes. Although this method is fully equivalent to the above functional approach \cite{Padmanabhan}, its physical basis arises much more transparently in the actual calculation. Explicitly, we have 
\begin{equation}
V (r) = \int \frac{d^3k}{(2\pi)^3} \ e^{-i\vec{k}\cdot\vec{r}} \ i\mathcal{M}(p)|_{(0,\vec{k})},
\label{potential formula}
\end{equation}
where $\mathcal{M} (p)$ is the Feynman amplitude in momentum space, which we can obtain from applying Feynman rules to the appropriate order. The prescription about the time of interaction $t_i$ in the previous paragraph translates into the choice of representations for the fields and vacuum in this diagrammatic approach. Particularly, the amplitude $\mathcal{M}$ is to be evaluated on the vacuum $|0\rangle_1$, while the representation of $\phi(x)$ is taken to be $\phi_{2}$. This is because the mass-shift does not change the original vacuum state $|0\rangle_1$ and the time $t_i$ is after the mass-shift, thus resulting in the dynamics being governed by $H_2$. The next subsection will discuss the impact of this choice on the Feynman propagator, which is where the mass-shift manifests its condensate of Eq.~(\ref{vacuum structure}) in the Feynman amplitude.   

\subsection{Feynman Propagator}

The Feynman propagator is defined in terms of the time-ordered product
\begin{eqnarray}
\Delta_F(x-y) = {}_1\langle 0| \mathcal{T} \left[ \phi(x) \phi(y) \right] |0\rangle_{1}.
\label{time-ordered prod}
\end{eqnarray}
One might quickly remark that because $\phi(x)$ is assumed to be unchanged in the mass-shift, Eq.~(\ref{time-ordered prod}) should reduce to the usual Feynman propagator with the physical mass $m_1$. However, since the physical field is governed by the equation of motion derived from the Hamiltonian $H_2$, the physical mode expansion is given by $\phi_2$ in Eq.~(\ref{field expansion mass 2}). This mismatch between the field representation and the vacuum state is fundamentally a quantum field-theoretic feature that has no classical or quantum-mechanical analog. In fact, a similar issue is also encountered in deriving the propagators of finite-temperature quantum field theory under the framework of TFD\cite{Takahashi,Umezawa book, Das}, where the field representation at zero-temperature is preserved while the vacuum state changes with temperature. In our case, we obtain
\begin{eqnarray}
\nonumber
\Delta_F(x) &=& {}_1\langle 0| \Theta\left(x^0\right) \phi(x) \phi(0) + \Theta\left(- x^0\right) \phi(0) \phi(x) |0\rangle_{1} \\
\nonumber
&=& {}_1\langle 0| \Theta\left(x^0\right)  \sum_{\vec{k},\vec{k}'} \frac{e^{i \vec{k} \cdot \vec{x}}}{2V \sqrt{\omega_2 \omega_2'}} \left( a_{-,2}^\dagger a_{-,2}'^{\dagger} e^{i\omega_2 t} + a_{-,2}^\dagger a_{2}' e^{i\omega_2 t} + a_{2} a_{-,2}'^\dagger e^{-i\omega_2 t} \right. \\
\nonumber
& & \Big. + a_{2} a_{2}' e^{-i\omega_2 t}  \Big)  + \Theta\left(- x^0\right) \sum_{\vec{k},\vec{k}'} \frac{e^{i \vec{k} \cdot \vec{x}}}{2V \sqrt{\omega_2 \omega_2'}} \left( a_{-,2}'^\dagger a_{-,2}^{\dagger} e^{i\omega_2 t} + a_{-,2}'^\dagger a_{2} e^{-i\omega_2 t}  \right. \\
& & \left. + a'_{2} a_{-,2}^\dagger e^{i\omega_2 t} + a'_{2} a_{2} e^{-i\omega_2 t}  \right) |0\rangle_{1}.
\end{eqnarray}

Making use of the Bogolyubov transformation in Eq.~(\ref{bogolyubov transformation}), we notice that the only non-zero terms must be proportional to either ${}_1 \langle 0 | a_{-,1} a_1'^\dagger |0 \rangle_{1} $ or $ {}_1 \langle 0 | a'_{1} a_{-,1}^\dagger |0 \rangle_{1}  $. Hence, it follows in the continuum limit of Eq.~(\ref{continuum limit}) that
\begin{eqnarray}
\nonumber
\Delta_F(x) &=&  \frac{1}{(2\pi)^3} \int d^3k \frac{e^{i \vec{k} \cdot \vec{x}} \left(v+u\right) }{2\omega_2} \left[ \Theta\left(x^0\right)  \left( v e^{i \omega_2 t} + u e^{-i \omega_2 t} \right) \right. \\
& & \left. + \Theta\left(-x^0\right) \left( u e^{i \omega_2 t} + v e^{-i \omega_2 t} \right)  \right].
\label{propagator long form}
\end{eqnarray}
It is important to recognize that unlike the usual scalar Feynman propagator, where only states with positive (negative) energies are propagated forward (backward) in time, Eq.~(\ref{propagator long form}) shows that because of the non-trivial vacuum structure in Eq.~(\ref{vacuum structure}) after the mass shift, there is always a mixture of both positive and negative states propagating both forward and backward in time. To proceed, we employ the following representation of the Heaviside step function,
\begin{equation}
\Theta (t) = -\frac{1}{2\pi i} \int_{-\infty}^{\infty} \frac{e^{-i t z }}{z + i \epsilon} dz,
\label{heaviside representation}
\end{equation}
in order to combine the four exponential time-factors in Eq.~(\ref{propagator long form}) with the spatial one into a Fourier transformation in spacetime. Using Eq.~(\ref{heaviside representation}), we can obtain the momentum representation $\Delta_F (p)$ from Eq.~(\ref{propagator long form}) by re-arranging it into 
\begin{eqnarray}
\Delta_F(x) &=&  \frac{i}{(2\pi)^4} \int d^4p \ \frac{\omega_2}{\omega_1} \frac{e^{-i px}}{E^2 - |\vec{k} |^2 - m_2^2 + i\epsilon}.
\label{propagator feynman coordinate}
\end{eqnarray}
Therefore, the Feynman propagator in momentum space is straightforwardly given from the coordinate representation in Eq.~(\ref{propagator feynman coordinate}) by
\begin{eqnarray}
\Delta_F(p)&=& \frac{\omega_2}{\omega_1} \frac{i}{p^2 - m_2^2 + i\epsilon}.
\label{feynman propagator momentum space}
\end{eqnarray}
In the limit $m_1 \rightarrow m_2$ we see that this reduces to the usual Feynman propagator with the physical mass $m_2$. Furthermore, perhaps we can interpret Eq.~(\ref{propagator feynman coordinate}) as that the effect of the condensate in Eq.~(\ref{vacuum structure}) due to the mass shift from $m_1$ to $m_2$ is to expand or contract the momentum volume measured by 
\begin{eqnarray}
d^4p \rightarrow \frac{\omega_2}{\omega_1} d^4p.
\end{eqnarray}
However, this deformation of the phase space is clearly suppressed out when $k \ll m_1, m_2$, yielding the usual Feynman propagator with the physical mass $m_2$. Hence, it is possible that high-energy experiments probing the condensate in Eq.~({\ref{vacuum structure}}) via the propagator of $\phi(x)$ might be not be sensitive enough to see these deviations. It is also important to note that the Feynman propagator in Eq.~(\ref{feynman propagator momentum space}) is no longer Lorentz-invariant because of the explicit dependence on the magnitude of the 3-momentum $k$ in $\omega_1$ and $\omega_2$. 

\subsection{The One-Boson Exchange Potential}

Using Feynman rules generated from the interaction Lagrangian density in Eq.~(\ref{L int density}), the Feynman amplitude is given by
\begin{equation}
\mathcal{M} = \frac{-i g_1 g_2}{p^2 - m_2^2} \frac{\omega_2}{\omega_1}.
\end{equation} 
The OBEP can then be computed straightforwardly from Eq.~(\ref{potential formula}) as the following Fourier transformation
\begin{eqnarray}
\nonumber
V (r) &=& - g_1 g_2\int \frac{d^3k}{(2\pi)^3} \  \frac{\omega_2}{\omega_1} \frac{e^{-i \vec{k} \cdot \vec{r}}}{ k^2 + m_2^2}\\
&=&-\frac{2g_1 g_2}{(2\pi)^2 r} \int_{0}^{\infty} dk \frac{k \sin kr}{ \sqrt{k^2 + m_1^2} \sqrt{k^2 + m_2^2}}.
\label{OBEP integral}
\end{eqnarray}
An interesting feature of Eq.~(\ref{OBEP integral}) is that it is symmetric in terms of $m_1$ and $m_2$. Although, the integral in Eq.~(\ref{OBEP integral}) does not  appear to have a closed form expression for arbitrary values of $m_1$ and $m_2$, we can study its behavior in several important special cases. 

\subsubsection{$m_1 =0$ or $m_2=0$}

\begin{figure}
	\centering
	\includegraphics[width=10cm]{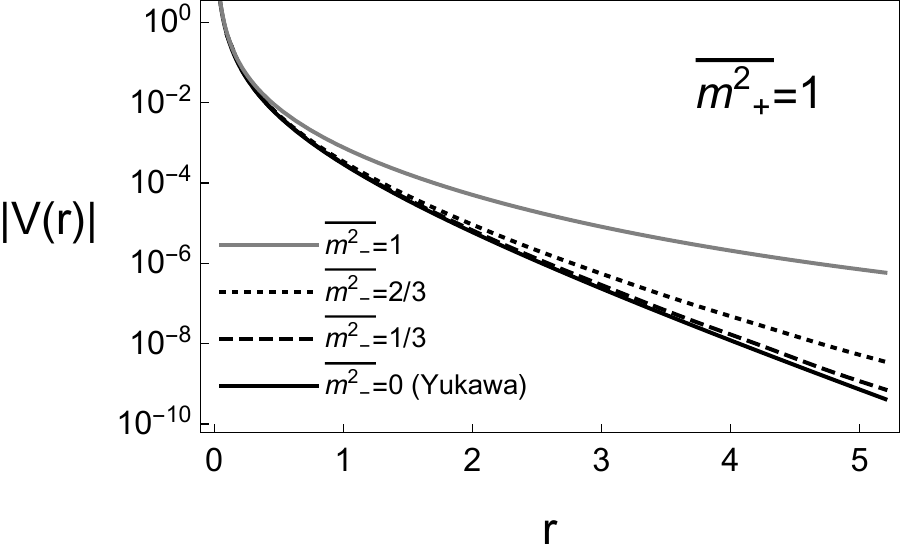} 
	\caption{Plot of the One-Boson Exchange Potential (in arbitrary units)  mediated by a scalar field after a mass-shift using Eq.~(\ref{OBEP integral}) assuming $\overline{m^2_+} = 1$, while various values of the mean difference of mass-squared ($\overline{m^2_-}$) are shown: $m_{2} =1$ [solid gray, which corresponds to the special case $m_{2} = 0$ given by Eq.~(\ref{V m2 = 0})], $m_{2} = 2/3$ (dotted black), $m_{2} =1/3$ (dashed black) and $m_{2} =0$ (solid black, which corresponds to the usual Yukawa potential with a boson mass $m_{1}=1$).}
	\label{fig OBEP}
\end{figure}

Taking without the loss of generality that $m_2=0$, the integral can then be evaluated exactly to yield
\begin{eqnarray}
\nonumber
V(r)  &=& -\frac{2g_1 g_2}{(2\pi)^2 r} \int_{0}^{\infty} dk \frac{\sin kr}{\sqrt{k^2 + m_1^2}}  \\
&=&   -g_1 g_2\left[ \frac{I_0 \left(m_1 r\right) - L_0 \left( m_1 r \right)}{4\pi r}\right],
\label{V m2 = 0}
\end{eqnarray}
where $I_0$ is the modified Bessel function of the first kind and $L_0$ is the modified Struve function.  A plot of Eq.~(\ref{V m2 = 0}) is shown by the gray solid line in Fig.~\ref{fig OBEP}. For short ranges,  i.e., $m_{1}r \ll 1$, Eq.~(\ref{V m2 = 0}) simplifies to the usual massless scalar potential
\begin{equation}
V(r) \simeq -\frac{g_{1}g_{2}}{4\pi r}.
\end{equation}

\subsubsection{$m_1,m_2 \neq 0$}

To study the general case, we first introduce some useful notation for the mean sum of mass-squared ($\overline{m^2_+}$) and the mean difference of mass-squared ($\overline{m^2_-}$):
\begin{equation}
\overline{m_{\pm}^2} \equiv \frac{m_1^2 \pm m_2^2}{2}.
\end{equation}
Then, the integral in Eq.~(\ref{OBEP integral}) can be evaluated by expanding the denominator of the integrand in a power series as follows
\begin{eqnarray}
\frac{1}{ \sqrt{k^2 + m_1^2} \sqrt{k^2 + m_2^2}} &=& \frac{1}{\sqrt{\left(k^{2} + \overline{m_{+}^{2}}\right)^{2} - \left(\overline{m_{-}^{2}}\right)^{2}}}
\nonumber \\
&=&   \sum_{n=0}^{\infty} \left(-1\right)^n \left( \begin{array}{c} 2n \\ n \end{array} \right) \left( \frac{ \overline{m_-^2}}{2}\right)^{2n} \frac{  \partial^{2n}_{\overline{m_+^2}} }{2n!} \left[ \frac{1 }{k^2 + \overline{m_+^2} } \right].
\end{eqnarray}
We note that this expansion is valid for all values of $k$ (uniformly convergent) as long as $\overline{m^2_+} > \overline{m^2_-}$, which is always true here since we are assuming $m_1,m_2 \neq 0$. 

It is then straightforward to obtain the OBEP when $m_1, m_2 \neq 0$ as follows
\begin{eqnarray}
\nonumber
V(r) &=& -\frac{g_1 g_2}{4\pi r} \sum_{n=0}^{\infty} \left( \begin{array}{c} 2n \\ n \end{array} \right) \left( \frac{ \overline{m_-^2} }{2}\right)^{2n} \frac{  \partial^{2n}_{\overline{m_+^2}} }{2n!} \left[ e^{- \sqrt{\overline{m_+^2}} r} \right] \\
&=& -\frac{g_1 g_2e^{- \sqrt{\overline{m_+^2}} r}}{4\pi r} + g_1 g_2\sqrt{\overline{m_+^2}}  \left( 1+ \sqrt{\overline{m_+^2}} r \right) \frac{ e^{- \sqrt{\overline{m_+^2}} r}}{64 \pi} \left( \frac{\overline{m_-^2}}{\overline{m_+^2}} \right)^2 
\nonumber \\
&&  \mbox{} + \mathcal{O} \left[ \left(\frac{\overline{m_-^2}}{\overline{m_+^2}} \right)^4 \right],
\label{OBEP final}
\end{eqnarray}
where it is clear that the OBEP's range of interaction after a mass-shift is characterized by $\overline{m^2_+}$, while the ratio $\overline{m^2_-}/\overline{m^2_+}$ signalizes its deviation from the  Yukawa potential in Eq.~(\ref{Yukawa Potential}). Since the mass of the boson characterizes the range of the interaction,  this deviation is only prominent at longer ranges as shown in Fig.~\ref{fig OBEP}; when $r \ll 1/\sqrt{\overline{m_{+}^{2}}}$, the potential reduces to the massless Yukawa potential result. Furthermore, we see that the effect of the condensate induced by the mass-shift is always an additional repulsive exponential term regardless of the values of $m_1$ and $m_2$.

\section{Casimir Effect}
\label{Casimir section}

In 1948, Casimir showed that there is a small attractive force between two parallel, uncharged, perfectly conducting plates \cite{Casimir}. Being a consequence of the electromagnetic interaction, it was at the time surprising that the force did not depend on the coupling constant of the interaction. Although it was pointed out later that the Casimir force is a special limit of the van der Waals interaction when retardation is included \cite{Casimir Polder}, the force can instead be interpreted as the quantum vacuum's reaction to  the boundary condition imposed by the plates \cite{Plunien}. It is important for us to note that while this formulation seemingly leads to the disappearance of the coupling constant in our setup, the coupling constant never truly disappears, it is only hidden in the prescription of boundary conditions on the field. In this section, we will examine effects of the condensate induced by the mass-shift in Eq.~(\ref{vacuum structure}) on the Casimir force due to the interaction with $\phi(x)$. Conceptually, our calculation of the Casimir force follows the approach  outlined previously by Mobassem \cite{Mobassem}.

Let us consider the scenario where we have two large, infinitely-thin, parallel plates at $\vec{l}_1 =(-L/2,0,0)$ and $\vec{l}_2 = (L/2,0,0)$ with distance $L$ apart from each other. These plates are coupled to our scalar field $\phi(x)$ via an interaction such that the free field is forced  to vanish at the two plates. Such boundary condition can be formalized as follows 
\begin{equation}
\phi(\vec{r}) = \begin{cases}
\displaystyle
\sum_{\vec{k}} \frac{1}{\sqrt{2V\omega_2}} \left[  a_2^{}  + a_{-,2}^\dagger   \right] e^{i \vec{k}_{||} \cdot \vec{r}_{||} }  \left( e^{i k_x x} - e^{-i k_x \left(x + L\right)} \right),   & x<-\frac{L}{2},  \\
\displaystyle
\begin{split}
&\sum_{\vec{k}_{||},n} \frac{1}{\sqrt{2V}} \left[ \frac{ a_2^{-} +    a_{-,2}^{-\dagger} } {\sqrt{\omega_2^-}} \sqrt{\frac{2}{L}} \sin\left(k^-_{x,n} x\right)    \right. \\
& \left. \mbox{} \,\,\,\,\,\,+  \frac{ a_2^{+} +  a_{-,2}^{+\dagger}}{\sqrt{\omega_2^+}} \sqrt{\frac{2}{L}} \cos\left( k_{x,n}^+ x\right)  \right] e^{i \vec{k}_{||} \cdot \vec{r}_{||} },  
\end{split}  & -\frac{L}{2} \leq x \leq \frac{L}{2}, \\
\displaystyle
\sum_{\vec{k}} \frac{1}{\sqrt{2V\omega_2}} \left[  a_2^{}  + a_{-,2}^\dagger   \right] e^{i \vec{k}_{||} \cdot \vec{r}_{||} }  \left( e^{i k_x x} - e^{-i k_x \left(x - L\right)} \right),   & x > \frac{L}{2}, \\
\end{cases}
\label{field casmir}
\end{equation}
where we define
\begin{eqnarray}
k_{x,n}^- &\equiv & \frac{2 n \pi}{L},\\
k_{x,n}^+ &\equiv & \frac{\left(2 n - 1\right) \pi}{L}.
\end{eqnarray}
Because we are considering our setup to be after the mass-shift, the $\phi_2$ representation is chosen. We also note that quantization on the newly prescribed boundary condition does not affect the Bogolyubov transformation in Eq.~(\ref{bogolyubov transformation}). 

The pressure from the vacuum at any point $\vec{r}$ of $\phi(x)$ can be extracted from the stress-energy tensor $T_{\alpha \beta}(x)$. For our massive scalar field, after the mass-shift $\phi(x)$ has the physical mass $m_2$ in the Lagrangian density, thus $T_{\alpha \beta}(x)$ is given by
\begin{eqnarray}
T_{\alpha \beta} = -\frac{\partial \phi}{\partial x ^\alpha} \frac{\partial \phi}{\partial x^\beta} + \frac{1}{2} \delta_{\alpha \beta} \left[ \vec{\nabla} \phi(x) \cdot \vec{\nabla} \phi(x) - \left( \frac{\partial \phi(x)}{\partial t} \right)^2 + m_2^2\phi^2(x)   \right]
\end{eqnarray}
Because we wish to calculate the Casimir pressure on the plates, due to the symmetry of the set-up, we only need to calculate the net pressure at the plate at $\vec{r}=\vec{l}_2$. This is given by the sum of pressures from the fields both inside and outside of the two plates
\begin{eqnarray}
\nonumber
P_{net} (x=L/2) &=& {}_1  \langle 0| \left[ \lim\limits_{x \rightarrow \left(L/2\right)^+} T_{11} \left(x\right) - \lim\limits_{x \rightarrow \left(L/2\right)^-}  T_{11} \left(x\right) \right] |0\rangle_1 \\
\nonumber
&=& \frac{\pi^2}{2VL}  \sum_{\vec{k}_{||},n} \left[ \frac{\left(2n-1\right)^2}{L^2} \frac{\left( u^+ + v^+\right)^2}{\omega_{2}^+} + \frac{\left(2n\right)^2}{L^2} \frac{\left( u^- + v^-\right)^2}{\omega_{2}^-} \right]\\
& &-\frac{1}{V} \sum_{\vec{k}} \frac{k_x^2}{\omega_{2}} \left(u+v\right)^2,
\label{pressure from stress tensor}
\end{eqnarray}
where we also have projected the net pressure on the $x$-axis.
We note that the first sum can be greatly simplified by observing that each term corresponds to either an odd ($2n-1$) or even ($2n$) term, which motivates the redefinition of the summation index to be $s$ such that
\begin{equation}
P_{net} (x=L/2) =  \frac{\pi^2}{2VL}  \sum_{\vec{k}_{||},s}  \frac{s^2}{L^2} \frac{\left( u + v \right)^2}{\omega_{2}} - \frac{1}{V} \sum_{\vec{k}} \frac{k_x^2}{\omega_{2}} \left(u+v\right)^2. 
\end{equation}
Then, after using the Bogolyubov transformation in Eq.~(\ref{bogolyubov transformation}), we obtain
\begin{eqnarray}
P_{net} (x=L/2) &=& \frac{\pi^2}{2VL^3}  \sum_{\vec{k}_{||},s} \frac{s^2}{\omega_1} - \frac{1}{V} \sum_{\vec{k}} \frac{k_x^2}{\omega_1}.
\end{eqnarray}
Finally, we use the following continuum limit for the summation
\begin{equation}
\frac{1}{V}  \sum_{\vec{k}_{||},s} \rightarrow \frac{1}{\left(2\pi\right)^2} \int d^2k_{||} \sum_{s},
\end{equation}
together with Eq.~(\ref{continuum limit}), and the substitution $ k_x = s\pi/L $ in the integration to obtain 
\begin{eqnarray}
P_{net} (x=L/2) &=& \frac{1}{\left(2\pi\right)^2} \int d^2k_{||} \left[ \frac{\pi^2}{2L^3} \left(  \sum_{s} \frac{s^2}{\omega_1} - \int_{0}^{\infty} ds\ \frac{s^2}{\omega_1} \right)  \right].
\label{casimir final integral}
\end{eqnarray}
Following Mobassem's use of the Poisson's sum formula, the integral Eq.~(\ref{casimir final integral}) can be evaluated to yield the net Casimir pressure as
\begin{equation}
P_{net} (x=L/2) = - \frac{m_{1}^4}{2 \pi^2} \sum_{s = 1}^{\infty} \frac{1}{2 m_1 L s} \left[ K_3 (2m_1 L s) - \frac{1}{2 m_1 L s} K_2(2m_1 L s) \right],
\label{casimir final} 
\end{equation}
which is rather surprising since it is the exact Casimir pressure \cite{Mobassem,Plunien} for a massive scalar field with mass $m_1$ in its free Lagrangian density. The mass-shift leaves no trace in the Casimir force, which suggests that somehow it does not care about the mass in the Lagrangian density. Indeed, this calculation shows that the Casimir force is only sensitive to the mass of the vacuum state. 

Eq.~(\ref{casimir final}) suggests that there exists a fundamental difference between a vacuum process as in the Casimir effect and the virtual-exchange as in the OBEP discussed in Section \ref{OBEP section}. A possible explanation for this behavior can be seen from the formalisms that we are using. Firstly, we see that in the OBEP calculation, the virtual-exchange is, in a sense, dynamical. The Feynman propagator propagates the field in time, whose evolution is dictated by the Hamiltonian $H_2$, derived from the Lagrangian density $\mathcal{L}_2$ and thus the dependence on the mass $m_2$ is eminent. Meanwhile, Eq.~(\ref{pressure from stress tensor}) clearly shows that Casimir force depends only on the local value of the field $\phi(x)$ at the boundaries. Hence, to a certain extent, the Casimir force is rather static in nature, thus $m_2$ does not appear at the end. Nonetheless, we want to re-iterate that the Casimir force as interpreted in our formulation has hidden its dependence on dynamical quantities such as the coupling constant and the mass $m_2$ in the interaction which is imposing boundary condition on the plates. Therefore, we believe that the result obtained here showing the Casimir force after a mass-shift to be independent of the mass after mass-shift must also be a certain limiting case of a more complicated situation.

\section{Conclusion}

We have considered here an effective mass-shift in the Lagrangian density of a scalar boson. In our model, the mass-shift leaves the vacuum state unchanged  while the field takes on an unitarily inequivalent representation of the new mass. Observers after the mass-shift thus perceive a condensate of pair-particles, which resembles the vacuum structure in an expanding universe \cite{Parker}. We then studied the consequences of this mass-shift in long-range forces. We showed that the condensate creates an repulsive exponential-decay correction in the OBEP compared to the usual Yukawa potential. The deviation is most notable in the long-range regime. Interestingly, we also showed that the mass-shift does not affect the Casimir pressure between two ideal parallel plates, indicating that the Casimir effect is only sensitive to the vacuum mass before the mass-shift.

Since the physics of phase-transitions in the early universe contains a rich variety of critical phenomena \cite{Boyanovsky}, it is difficult to be certain at the moment about the applicability of our model to particles in the SM. Phenomenologically speaking, there exist 2 other non-trivial possibilities, which were not considered in this work. Firstly, the mass-shift might manifest as a change in the vacuum state, while the physical representation of the field remains. An example of such manifestation is the TFD formulation of finite-temperature QFT \cite{Umezawa book,Das,Takahashi}. Because our results for the OBEP in Section \ref{OBEP section} are conditioned only upon the mismatch between the representations of the field and the vacuum, we expect that such results still hold. However, our results for the Casimir force in Section \ref{Casimir section} would change since it was shown to depend explicitly on the vacuum. Secondly, it is also possible that after the abrupt mass-shift, the vacuum slowly transitions to and eventually settles down at $|0\rangle_2$. Hence, we expect our result will hold for a brief period of time after the mass-shift, but not in measurements carried out today. Nonetheless, such phenomenon effectively behaves as if the strength of the OBEP varies through time and reaches its maximum in our time. Should such variation be sufficiently large, it can leave certain remnants in early structure formation and other cosmological observations \cite{Uzan}. While it is currently not clear which model best describes the Higgs phase transition in the SM, we emphasize  that our model might be relevant to BSM physics, particularly those hidden in the dark sector.

Our results are certainly meaningful for the search for new physics via long-range forces and potentials \cite{review new physics atoms, review new physics macroscopic}. The results derived here stem from new physics that resides in the non-trivial vacuum structure. Particularly, we dichotomized the low-energy signature of the mass-shift in the OBEP and the Casimir force. While we pointed out that this sharp distinction should only be valid within certain limits, it would be interesting to further explore both theoretically and experimentally this relationship in future study. Furthermore, with the employment of unitarily inequivalent vacua, we hope to have demonstrated its potential as a new tool for future phenomenological work that explore the mysteries of BSM physics.

\section*{Acknowledgments}

We would like to express our gratitude to the two anonymous reviewers for pointing out seminal references related to our work and suggesting further discussion on our results.

\end{document}